\documentclass[aps,twocolumn,showpacs,preprintnumbers,amsmath,amssymb]{revtex4}

\usepackage{epsfig}
\usepackage{enumerate}

\newcommand{\eqsref}[1]{(\ref{#1})}
\newcommand{\med}[1]{\langle #1 \rangle}

\begin{document}

\preprint{APS/2-HEP}

\title{Transverse Momentum Fluctuations \\ 
from \\
Clustering and Percolation of Strings} 

\author{J. Dias de Deus}
\author{A. Rodrigues} 

\altaffiliation[Also at ]{Escola Secund\'{a}ria da Ramada, Portugal.}
\affiliation{CENTRA and Departamento de F\'{\i}sica (IST),\\
Avenida Rovisco Pais, 1049-001 Lisboa, Portugal}

\date{\today}

\begin{abstract}
Transverse momentum fluctuations can be understood as resulting from
clustering of strings or partons. Data allows to distinguish 
clustering without percolation, from clustering with percolation.
Percolation is clearly favored by data.
\end{abstract}


\maketitle

Event by event fluctuations of total transverse momentum, $P$, in high energy 
hadronic and heavy ion collisions, may give relevant information on the
thermodynamical properties of dense hadronic matter, and, in particular, on
the existence of a phase transition \cite{ref:1,ref:2}. 
The problem has been
discussed in a series of experimental papers \cite{ref:3,ref:4,ref:5}.\\
In this paper we argue that the occurrence of $P$ fluctuations is not 
necessarily a signal of a thermal phase transition. It may rather be a 
geometrical effect, due to clustering and percolation of strings 
\cite{ref:6,ref:7}.\\
Let us suppose that in hadron-hadron and nucleus-nucleus collisions one always
produces a cluster of $N$ independent identical strings. For the $P$ 
distribution one can then write  
\begin{equation}
  f(P)_{N}=\int \Pi_{i}f(P_{i})_{1}dP_{i} \,\,\delta(P-\sum_{i=1}^{N}P_{i}). 
       \label{eq:1}
\end{equation}
From Eq. \eqsref{eq:1} it follows that
\begin{equation}
  \med{P}_{N} = N \med{P}_{1} 
       \label{eq:2}
\end{equation}
and
\begin{equation}
  \med{P^2}_{N} = N \med{P^2}_{1}+N(N-1)\med{P}_{1}^{2}, 
       \label{eq:3}
\end{equation}
such that
\begin{equation}
  \med{P^2}_{N}-\med{P}_{N}^{2} = N \left[ \med{P^2}_{1}-\med{P}_{1}^{2})\right], 
       \label{eq:4}
\end{equation}
as expected for independent strings.\\
However, if one considers fluctuations in the number $N$ of strings per cluster, 
one obtains
\begin{equation}
  \med{P^2}-\med{P}^{2} = \left[\med{N^2}-\med{N}^{2}\right]\med{P^2}_{1}+
  \med{N}\left[ \med{P^2}_{1}-\med{P}_{1}^{2}\right] 
       \label{eq:5}
\end{equation}
Note that if one assumes that particles from the single string are emitted 
independently (Poisson like distribution), one arrives at:
\begin{equation}
  \med{P^2}_{1}-\med{P}^{2}_{1} = \bar{n}\left(\med{p^2}-\med{p}^2 \right), 
       \label{eq:6}
\end{equation}
where $\bar{n}$ is the string average multiplicity and $p$ refers to the 
particle transverse momentum.\\
Introducing now the quantity $F_{P_{T}}$ (see, for instance, Refs. 
\cite{ref:3,ref:4}),
\begin{equation}
 F_{P_{T}}=\sqrt{ \frac{\med{Z^2}} {\med{n}\med{z^2}} }-1,
       \label{eq:7}
\end{equation}
with $\med{Z^2}=\med{P-\med{P}}^2$, $\med{z^2}=\med{p-\med{p}}^2$ and 
$\med{n}=\med{N}\bar{n}$, such that $F_{P_{T}}=0$ for random emission, we
obtain
\begin{equation}
 F_{P_{T}}=\sqrt{c\frac{\med{N}}{k}+1}-1,
       \label{eq:8}
\end{equation}
where
\begin{equation}
 \frac{1}{k}=\frac{\med{N^2}-\med{N}^2}{\med{N}^2}
       \label{eq:9}
\end{equation}
is the normalized variance of the cluster distribution, and
\begin{equation}
 c=\bar{n}\frac{\med{p}^2}{\med{p^2}-\med{p}^2}
       \label{eq:10}
\end{equation}
is characteristic of the string particle production mechanism. 
The inclusion of color summation factors when forming clusters of strings, not 
considered here, may affect the value of c, Eq. \eqsref{eq:10}.  
In any case, physics 
is mostly determined by $\med{N}/k$.\\
Experimentally, data on $F_{P_{T}}$, Ref. \cite{ref:4}, show that 
this quantity increases as matter density (number of participant nucleons) 
increases up to some point starting to decreases for larger densities. Making
use of two very simple string clustering models, Model I without percolation
and Model II with percolation, we show that only Model II, with percolation, 
is consistent with data. \\
Let us consider the (two dimensional) problem of distributing $N_{s}$ strings
with a definite transverse area, in a total area $M$ (the impact 
parameter transverse interaction area). A relevant parameter is the transverse 
area density $\eta$: 
\begin{equation}
 \eta=\frac{N_{s}}{M}.
       \label{eq:11}
\end{equation}\\
Let us now assume that in a collision 
clusters of strings are formed, $P(N)$ being the probability of having a
 $N$-cluster(cluster with $N$ strings).\\
In principle - by using Monte Carlo simulations - it is possible to calculate 
$\med{N_{c}}$, the average number of clusters, $\med{N}$ the average number 
of strings per cluster, higher moments of $P(N)$, and the average area 
occupied by the clusters $\med{A}$.\\
There are two sum-rules which have, in general, to be satisfied:
\begin{equation} 
	\text{i.}\qquad\qquad\qquad\qquad
		\med{N_{c}}\med{N}=N_{s},\qquad\qquad
		\label{eq:12} 
\end{equation}%
\begin{equation}  
	\text{ii.}\qquad\qquad\qquad
		\med{N_{c}}\med{A}=M(1-e^{-\eta}).\qquad       
		\label{eq:13} 
\end{equation}%
The first sum-rule expresses the conservation of number of strings
and the second is simply the area sum-rule: the right hand side of this sum-rule 
is the overall area occupied by the clusters, Refs. \cite{ref:8,ref:9}.
Note that Eq. \eqsref{eq:12} and Eq. \eqsref{eq:13} are invariants, independent
of percolation. Percolation affects $\med{A}$, $\med{N}$ and $\med{N}_{c}$, 
but not $N_s$ and $M(1-e^{-\eta})$.\\
In the next two sections we introduce the two Models that illustrate our arguments.

\subsection*{Model I - No percolation}
This model is essentially equivalent to the problem of distributing $N_s$ coins among $M$ 
boxes. The total area is $M$, each box has area 1, and $\med{A}=1$. From 
Eq. \eqsref{eq:12} and Eq. \eqsref{eq:13} one immediately obtains 
\begin{equation} 
	\med{N}=\frac{\eta}{1-e^{-\eta}}\;.   
	\label{eq:14} 
\end{equation}
The second moment $\med{N^2}$ of $P(N)$ can also be computed and we obtain
for $k$ the following equation,
\begin{equation} 
	k=\frac{\med{N}^2}{\med{N^2}-\med{N}^2}=\frac{\eta}{1-(1+\eta)e^{-\eta}}\;.   
	\label{eq:15} 
\end{equation}
Note that as $\eta\rightarrow 0$, $\med{N}\rightarrow 1$ and 
$k\rightarrow \infty$. This means that when the coin density is very low, the
clusters have just one coin and the distribution is a $\delta$-function 
($\med{N^2}-\med{N}^2\rightarrow 0$). In the $\eta\rightarrow\infty$ limit, 
$\med{N^2}-\med{N}^2\rightarrow \med{N}\sim\eta$, increases with 
$\eta$. The behavior of the quantity of interest to us, $\med{N}/k$, is show, 
in the case of Model I, in Fig.(1).

\begin{figure}
  \begin{center}
  \mbox{\epsfig{file=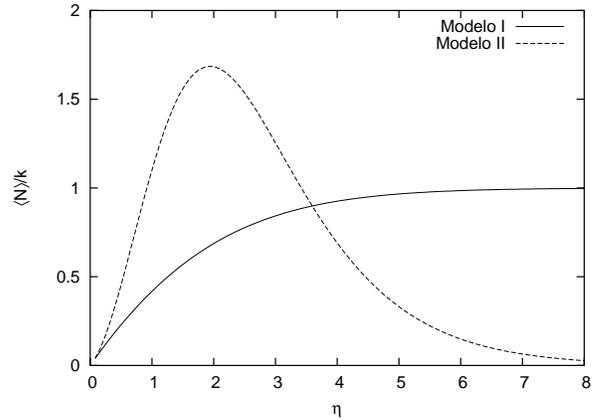,width=0.45\textwidth}}
  \end{center}
  \caption[Average strings]{Average number of strings per cluster multiple by 
  the normalize variance of the cluster distribution as function
  of dimension less density $\eta$ in the context of AA collisions geometry, 
  for Model I and Model II. In both Models $M=16$.}\label{fig:1}
\end{figure}

\subsection*{Model II - Percolation}
In the present context percolation means that the average area of the cluster
can not be constant: it must increase with density. In the low density limit, 
we can accept that Model I is correct, i.e.,  
\begin{equation} 
	\med{A} \stackrel{_{\eta\rightarrow 0}}{\longrightarrow} 1. 
	\label{eq:16} 
\end{equation}
However, in the $\eta\rightarrow\infty$ limit, percolation requires
\begin{equation} 
	\med{A} \stackrel{_{\eta\rightarrow\infty}}{\longrightarrow} 
	M(1-e^{-\eta}),
	\label{eq:17} 
\end{equation}
or, $\med{N_c}\rightarrow1$. In this case, $P(N)$ is dominated by big 
size clusters. Percolation implies that
we have a $\delta$-function distribution, not only in the $\eta\rightarrow 0$ 
limit, $P(N)=\delta(N-1)$, but a $\delta$-function distribution, as well, in the
$\eta\rightarrow \infty$ limit, $P(N)\rightarrow\delta(N-\med{N})$.\\
In order to satisfy Eq. \eqsref{eq:16} and Eq. \eqsref{eq:17} we choose for 
$\med{A}$, as a function of $\eta$, the function 
\begin{figure}
  \begin{center}
  \mbox{\epsfig{file=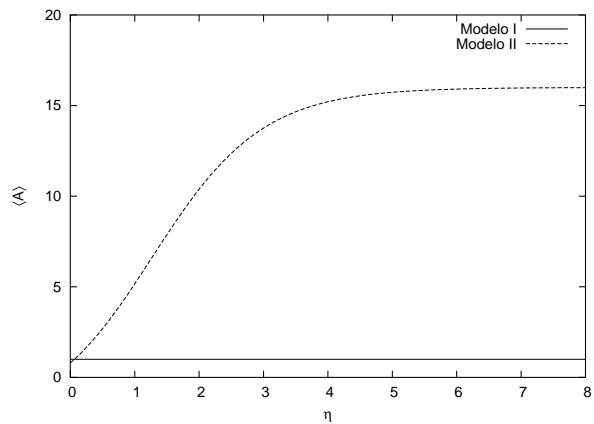,width=0.45\textwidth}}
  \end{center}
  \caption[Average area]{Avarage area occupied by a cluster as function
  of dimension less density $\eta$ for AA collisions geometry, for 
  Model I and Model II. In both Models $M=16$.}\label{fig:2}
\end{figure}
\begin{equation} 
	\med{A}=f(\eta)\left[M(1-e^{-\eta})-1\right]+1  
	\label{eq:18} 
\end{equation}
where $f(\eta)$ represents a percolation function, $f(\eta)=0$ means no percolation,
$\med{A}=1$, $f(\eta)=1$ means full percolation,
with the Heavyside function approximated by a smoother function, because of the 
finite size of the systems in simulation,(see last paper in Ref. \cite{ref:6})
\begin{equation} 
	f(\eta)=\left(1+e^{-(\eta-\eta_c)/a} \right)^{-1}  
	\label{eq:18a} 
\end{equation}
with this function we fit the fraction of occupied area by all string in the transverse plane
and we obtain for $a=0.85\pm 0.01$ and $\eta_c=1.155\pm 0.015$ this parameter 
is the percolation threshold. \\
In Fig.(2) we show $\med{A}$ as a function, of $\eta$ for Model I and II. From
Eqs. \eqsref{eq:12}, \eqsref{eq:13} and \eqsref{eq:18} one immediately 
obtains
\begin{equation} 
	\med{N}=\frac{\eta}{1-e^{-\eta}}(f(\eta)\left[M(1-e^{-\eta})-1\right]+1)  
	\label{eq:19} 
\end{equation}
with limits $\med{N}\rightarrow 1$ as $\eta\rightarrow 0$, as in Eq. \eqref{eq:14}, 
and $\med{N}\rightarrow N_{s}$, as $\eta\rightarrow\infty$ (the average 
cluster, asymptotically, contains all the strings).\\
In order to estimate the second moment $\med{N^2}$, or the variance, one
has to keep in mind that:
\begin{equation*} 
	\med{N^2}-\med{N}^2 \geq 0\;,
\end{equation*}
\begin{equation} 
		\med{N^2}-\med{N}^2 \stackrel{_{\eta\rightarrow 0}}{\longrightarrow} 0\;, 
	\label{eq:20} 
\end{equation}
\begin{equation*} 
		\med{N^2}-\med{N}^2 \stackrel{_{\eta\rightarrow \infty}}{\longrightarrow} 0\;.
\end{equation*}
By making use of Eqs. \eqsref{eq:20} we write
\begin{equation} 
	\med{N^2}=\frac{1-e^{\left[-(1+\eta)(1-\exp(-\eta))\right]}}
	  {1-e^{-\eta}}\,\, \med{N}^2\;.
	\label{eq:21} 
\end{equation}
It is easily seen that Eq. \eqsref{eq:21} satisfies the constraints of 
Eqs. \eqsref{eq:20}.\\
In Fig.(1) we also show the quantity $\med{N}/k$ for Model II. It is clear 
what is the essential difference between Model I (No Percolation) and 
Model II (Percolation). While in Model I, $\med{N^2}-\med{N}^2$ always increases
with $\eta$, thus originating increasing $P$ fluctuations, in Model II, as
$\eta\rightarrow\infty$, $\med{N^2}-\med{N}^2$ starts to decrease, thus 
originating decreasing $P$ fluctuations.

\begin{figure}
  \begin{center}
  \mbox{\epsfig{file=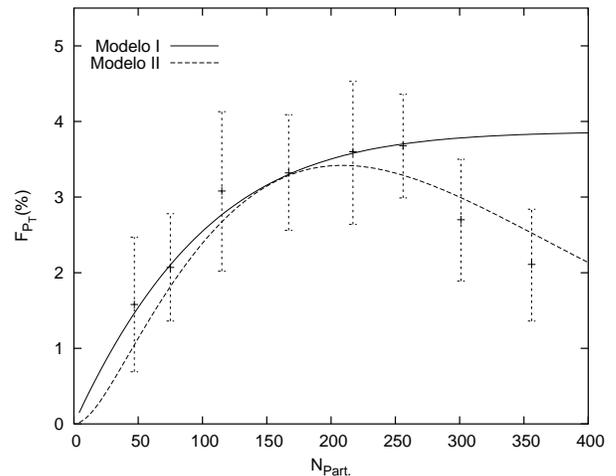,width=0.45\textwidth}}
  \end{center}
  \caption[Models and data]{Data from PHENIX experiment compared 
  with Model I($M=220$, $c=0.220$, $N_{s}^{p}=8$) and Model II($M=420$, $c=0.430$, $N_{s}^{p}=8$) 
as a function of $N_{part}$.}\label{fig:3}
\end{figure}

\subsection*{Data compared with models}
Before comparing Models I and II with data, by making use of Eq. \eqsref{eq:8},
we would like to give estimates for the values of $M$ and $c$.\\
In string percolation the variable $\eta$, Eq. \eqsref{eq:11}, is defined as
\begin{equation} 
	\eta=\left(\frac{r}{R}\right)^2N_{s},
	\label{eq:22} 
\end{equation}
where $r$ is the transverse radius of the string (we assume $r=0.2$ fm) and $R$
is the average interaction radius. It is natural to identify $M$, which has 
the meaning of an area, with $(R/r)^2$. We then have
\begin{equation} 
	M\leq\left(\frac{R_{A}}{r}\right)^2,
	\label{eq:23} 
\end{equation}
where $R_{A}$ is the radius of the nucleus with, for gold, $R_{A}=6.6$ fm. We thus 
obtain the bounds
\begin{equation} 
	25\leq M\leq 1100,
	\label{eq:24} 
\end{equation}
where the left hand side represents the limit of pp collision.
Concerning the parameter $c$, Eq. \eqsref{eq:10}, the quantity 
$\med{p}^2/(\med{p^2}-\med{p}^2)=1/(1-\pi/4)$, if we admit, as in 
Schwinger model, that the particle $p_{T}$ distribution of the string is 
gaussian. The quantity $\bar{n}$, the string particle density, was in 
Ref. \cite{ref:10} estimated to be $0.7$, for valence strings, and $20$ times 
smaller for sea strings. We obtain the bounds,
\begin{equation} 
	0.163\leq c\leq 3.262
	\label{eq:25} 
\end{equation}
Note that $c$ should be a decreasing function of energy, as the fraction of 
sea strings increases with energy.\\
The number of produced strings is obtained from the number of participants
by making use of the approximate usual formula
\begin{equation} 
	N_{s}=N_{s}^{p}N_{A}^{4/3},
	\label{eq:26} 
\end{equation}
where $N_{s}^{p}$ is the average number of strings in pp collision, 
$N_{s}^{p}$ is assumed to be equal $8$ at $\sqrt{s}=200$ GeV, which agree with Ref. \cite{ref:11}, and
$N_{A}=N_{part}/2$.\\
In Fig.(3) we compare our Models I and II with data published in 
Ref. \cite{ref:12}. The parameters $M$ and $c$ were, within bounds, inequalities 
Eq. \eqsref{eq:24} and Eq. \eqsref{eq:25}, adjusted separatly for each one 
of the models. \\
In conclusion, data clearly favour Model II, with percolation.\\
We would like to thank Elena Ferreiro, Carlos Pajares and Roberto Ugoccioni
for advice and discussions.

\newpage
 


\begin{thebibliography}{10}
\bibitem{ref:1}                 
L. Stodolsky, Phys. Rev. Lett. {\bf 75}, 1044 (1995);\\
E. Shuryak, Phys. Lett. B {\bf 439}, 9 (1998);\\
M. Stephanov, K. Rajagopal and E. Shuryak, Phys. Rev. D {\bf 60}, 114028 (1999).

\bibitem{ref:2}                 
M. Gazdzicki and S. Mr\'{o}wczynski, Z. Phys. C {\bf 54}, 127 (1992);\\
R. Rybczynski, Z. Wlodarczyk and G. Wilk, hep-ph/0305329.

\bibitem{ref:3}                 
M. Tannenbaum, Phys. Lett. B {\bf 498}, 24 (2001);\\
H. Appelshauser {\em et al.}, NA49 Collaboration, Phys. Lett. B {\bf 459}, 679 (1999);\\
H. Appelshauser {\em et al.}, CERES Collaboration, Nucl. Phys. A {\bf 698}, 253c (2002).

\bibitem{ref:4}                 
K. Adcox {\em et al.}, PHENIX Collaboration, Phys. Rev. C {\bf 66}, 024901 (2002);\\
J. Nystrand, Nucl. Phys. A {\bf 715}, 603c (2003)

\bibitem{ref:5}                 
R. L. Ray {\em et al.}, STAR Collaboration, Nucl. Phys. A {\bf 715}, 45c (2003).

\bibitem{ref:6}                 
N. Armesto, M. A. Braun, E. G. Ferreiro and C. Pajares, Phys. Rev. Lett. {\bf 77}, 3736 (1996); \\
M. Nardi and H. Satz, Phys. Lett. B {\bf 442}, 14 (1998);\\
J. Dias de Deus, R. Ugoccioni and A. Rodrigues, Eur. Phys. J. C {\bf 16}, 537 (1996).

\bibitem{ref:7}                 
E. G. Ferreiro, F. del Moral and C. Pajares, hep-ph/0303137. 

\bibitem{ref:8}                 
D. Stauffer, Phys. Rep. {\bf 54}, 2 (1979).

\bibitem{ref:9}                 
M. A. Braun, C. Pajares and R. Ranft, Int. J. Mod. Phys. A {\bf 14}, 2689 (1999);

\bibitem{ref:10}                
J. Dias de Deus and R. Ugoccioni,  Phys. Lett. B {\bf 491}, 253 (2000).

\bibitem{ref:11}                
N. S. Amelin, M. A. Braun and C. Pajares, Z. Phys. C {\bf 63}, 507 (1994).

\bibitem{ref:12}                
J. Nystrand, PHENIX Collaborattion,  Nucl. Phys. A {\bf 715}, 603c (2003).

\end{thebibliography}
\end{document}